\documentclass[12pt]{iopart}

\usepackage{dcolumn}
\usepackage{bm}
\usepackage{ifpdf}
\usepackage{hyperref}
\usepackage{dcolumn}
\usepackage{bm}
\usepackage[spanish,english]{babel}
\usepackage{amsfonts}
\usepackage{amssymb}
\usepackage{graphicx}
\usepackage[latin1]{inputenc}

\newcommand{\be}{\begin{equation}}
\newcommand{\en}{\end{equation}}
\newcommand{\bea}{\begin{eqnarray}}
\newcommand{\ena}{\end{eqnarray}}
\newcommand{\bes}{\begin{subequations}}
\newcommand{\ees}{\end{subequations}}

\begin{document}

\title{Geodesically complete BTZ-type solutions of $2+1$ Born-Infeld gravity}
\author{D. Bazeia$^1$, L. Losano$^1$, Gonzalo J. Olmo$^{2,1}$, D. Rubiera-Garcia$^3$}
\address{$^1$Departamento de F\'isica, Universidade Federal da
Para\'\i ba, 58051-900 Jo\~ao Pessoa, Para\'\i ba, Brazil}
\address{$^2$Departamento de F\'{i}sica Te\'{o}rica and IFIC, Centro Mixto Universidad de Valencia - CSIC.
Universidad de Valencia, Burjassot-46100, Valencia, Spain}
\address{$^3$Instituto de Astrof\'isica e Ci\^encias do Espa\c{c}o, Universidade de Lisboa, Faculdade de Ci\^encias, Campo Grande, PT1749-016 Lisboa, Portugal}

\eads{\mailto{bazeia@fisica.ufpb.br}, \mailto{losano@fisica.ufpb.br},  \mailto{gonzalo.olmo@uv.es}, \mailto{drgarcia@fc.ul.pt}}

\date{\today}
\begin{abstract}
We study Born-Infeld gravity coupled to a static, nonrotating electric field in $2+1$ dimensions and find exact analytical solutions. Two families of such solutions represent geodesically complete, and hence nonsingular, spacetimes. Another family represents a point-like charge with a singularity at the center. Despite the absence of rotation, these solutions resemble the charged, rotating BTZ solution of General Relativity but with a richer structure in terms of horizons. The nonsingular character of the first two families turn out to be attached to the emergence of a wormhole structure on their innermost region. This seems to be a generic prediction of extensions of General Relativity formulated in metric-affine (or Palatini) spaces, where metric and connection are regarded as independent degrees of freedom.
\end{abstract}

\pacs{04.20.Jb, 04.40.Nr, 04.50.Kd, 04.70.Bw}

\submitto{\CQG}

\maketitle

\section{Introduction}

The year 1992 awaken with the striking and unexpected finding by Ba\~nados, Teitelboim and Zanelli (BTZ) of a class of vacuum solutions of the $2+1$ dimensional Einstein field equations of General Relativity (GR) with a negative cosmological constant $\Lambda=-1/l_{\Lambda}^2<0$. In Schwarzschild-like coordinates, these solutions are given by the metric \cite{BTZ}

\begin{eqnarray}
ds^2&=&-N(r)dt^2+N(r)^{-1}dr^2+r^2(d\phi +N^{\phi}dt)^2 \nonumber \\
N(r)&=&-M+\frac{r^2}{l_{\Lambda}^2} + \frac{J^2}{4r^2} \hspace{0.2cm} ; \hspace{0.2cm} N^{\phi}=-\frac{J^2}{2r^2} \label{eq:BTZ}
\end{eqnarray}
and are characterized by mass, $M$, cosmological constant length squared, $l_{\Lambda}^2$, and angular momentum, $J$. The relevance of this geometry is twofold: first for $M>0$ it describes a family of black hole solutions with two (inner and event) horizons, provided that $M > \vert J \vert l_{\Lambda}^{-1}$ (when $M = \vert J \vert l_{\Lambda}^{-1}$ the two horizons merge into an extreme black hole), and sharing many physical features with the four dimensional Kerr black hole, a finding that seemed to contradict previous results in the literature by that time \cite{Ida}. In addition, the solution $M=-1$ (and $J=0$) emerges as a three dimensional Anti-de Sitter (AdS$_3$) state, disconnected from the spectra of black hole solutions with $M>0$ by a mass gap. Such a state contains no horizons, but there is no curvature singularity to hide, either, a finding that represented a great insight on the understanding of spacetime singularities in the context of classical gravitation. In the gap $-1<M<0$ separating the black hole solutions from the AdS$_3$ state, one finds naked singularities that can be interpreted as topological defects supported by a $0$-brane \cite{Miskovic}. The vacuum BTZ solution can be enlarged to include an electric charge $Q$, with similar nice properties \cite{Carlip:1995qv,MTZ}.

Though it took some time to fully understand and appreciate its features, along the years its impact upon the AdS/CFT correspondence \cite{AdSCFT,Navarro}, thermodynamic properties \cite{ThermoBTZ}, connections to solid state physics \cite{SSP} and regular solutions has significantly increased (see \cite{Birmingham01} for a detailed analysis of the properties of the BTZ solution). Indeed, it has triggered a great deal of interest upon GR in $2+1$ dimensions, in order to get further insights on the fundamental nature of gravitation using a scenario where the inherent mathematical complexity of GR is significantly softened. Among the problems studied in the $2+1$ context so far, we mention the magnetic counterparts of the electrically charged BTZ solution \cite{MagneticBTZ}, gravitational collapse \cite{Matschull}, geometric and thermodynamic features of several non-linear models \cite{NED3D}, wormholes \cite{WHs3D}, BTZ-like solutions out of the coupling to scalar fields \cite{Scalar}, hairy BTZ-like black holes \cite{Hairy}, or thin-shell solutions \cite{Lemos}.

A different viewpoint on this subject is to consider extensions of GR, which have been widely investigated over the last decade both on their theoretical and phenomenological aspects in connection with high-energy physics and cosmology (see e.g. \cite{Reviews} for some reviews on the subject). In this sense, BTZ-type solutions have been studied within $f(R)$ gravity \cite{fR3D}, dilaton gravity \cite{di}, teleparallel and $f(T)$ gravities \cite{tele}, Chern-Simons \cite{CS}, Kaluza-Klein \cite{KaKl}, noncommutative geometries \cite{NC} and Brans-Dicke theory \cite{BD}. Our work is placed in this context and its main aim is to find the counterpart of the electrically charged $2+1$-dimensional BTZ solution without rotation in the context of an interesting proposal for extending GR, dubbed Born-Infeld gravity \cite{BIgrav,BIgravb,Comelli:2005tn}. This theory has attracted a good deal of attention in the last few years due to its many applications regarding astrophysics \cite{BIastro}, black hole physics \cite{BIbh} and cosmology \cite{BIcosmo}. This is, in particular, due to the fact that the field equations of the theory have the appealing feature of being ghost-free and second-order when formulated in the Palatini approach, where metric and connection are regarded as independent entities (see \cite{or11} and \cite{Olmo2016} for a detailed description of this approach). This is in sharp contrast with the generic higher-order derivative field equations typically found in the metric formulation of modified gravity (where the affine connection is taken \emph{a priori} to be given by the Christoffel symbols of the metric). Note, however, that the teleparallel version of the theory also satisfies second-order equations \cite{Ferraro:2008ey}.

Recently it was found that the four dimensional Born-Infeld gravity coupled to electromagnetic fields supports geometries that are free of spacetime singularities for the full spectrum of mass and charge \cite{ors15v}. The last statement holds true regardless of the potential blow up of curvature scalars in some cases, whose meaning and impact were subsequently analyzed in Ref.\cite{ors16}. As will be shown here, in the $2+1$ dimensional electrovacuum scenario considered in this work, the corrections induced by Born-Infeld gravity have their reflection on the geometry in that, besides a cosmological constant term, a new term which is formally similar to the $J^2$ contribution of the rotating BTZ solution appears. Significant novelties as compared to that solution also involve a richer description in terms of horizons, and a non-trivial modification of the innermost region, where a wormhole structure arises for two families of solutions. This wormhole provides a geodesically complete and hence regular spacetime for all the spectrum of mass, charge, cosmological constant and Born-Infeld gravity length scale. In addition, another family of solutions without wormhole structure and a point-like singularity at the center is also discussed.

The content is organized as follows: in Sec.\ref{sec:II} we introduce the theory and the conventions, derive the field equations and solve them for a circularly symmetric electrostatic (Maxwell) field. The properties of the resulting solutions are discussed in Sec.\ref{sec:III}, where the analysis is split into two families of wormhole solutions and a non-wormhole one, paying special attention to the structure of horizons and discussing briefly the thermodynamics. The geodesic behaviour for all these families is considered in Sec.\ref{sec:Geo} and we conclude in Sec.\ref{sec:V} with a discussion and some perspectives.

\section{Action, field equations, and solutions} \label{sec:II}

The action of Born-Infeld gravity in $d$ spacetime dimensions can be written as
\begin{eqnarray} \label{eq:EiBI}
S_{BI}&=&\frac{1}{\kappa^2\epsilon}\int d^dx \left[\sqrt{-|g_{\mu\nu}+\epsilon R_{\mu\nu}(\Gamma)|}-\lambda \sqrt{-g}\right] \nonumber \\
&+& S_m(g_{\mu\nu},\psi_m),
\end{eqnarray}
with the following definitions: $\kappa^2\equiv 8\pi l_P^{d-2}$ defines the $d$-dimensional Newton's constant in terms of the Planck length $l_P$, vertical bars denote a determinant while $g$ is the determinant of the spacetime metric $g_{\mu\nu}$, which is a priori independent of the affine connection $\Gamma \equiv \Gamma_{\mu\nu}^{\lambda}$ (Palatini approach). The Ricci tensor $R_{\mu\nu}(\Gamma) \equiv {R^\alpha}_{\mu\alpha\nu}$, where ${R^\alpha}_{\beta\mu\nu}=\partial_{\mu} \Gamma^{\alpha}_{\nu\beta}-\partial_{\nu} \Gamma^{\alpha}_{\mu\beta}+\Gamma^{\alpha}_{\mu\lambda}\Gamma^{\lambda}_{\nu\beta}-\Gamma^{\alpha}_{\nu\lambda}\Gamma^{\lambda}_{\mu\beta}$ is the Riemann tensor, is entirely built out of the connection $\Gamma_{\mu\nu}^{\lambda}$. $S_m$ is the matter action, which is assumed to depend only on the metric and the matter fields, that are collectively labelled as $\psi_m$. The meaning of the (length squared) parameter $\epsilon$, and of the constant $\lambda$ in the action (\ref{eq:EiBI}), follows from an expansion in series of $\epsilon \ll 1$, which yields
\begin{eqnarray} \label{eq:action-low}
\lim_{\epsilon\to 0}S_{BI}&=&\int d^d x\sqrt{-g}\left[\frac{(1-\lambda)}{\epsilon\kappa^2}+\frac{1}{2\kappa^2}R \right] \nonumber \\
&+& \epsilon \int d^dx \sqrt{-g} \left[ \frac{1 }{4\kappa^2}  \left( \frac{1}{2} R^2-R_{\mu\nu}R^{\mu\nu}\right) \right] \\
&+& O(\epsilon^2) +  S_m. \nonumber
\end{eqnarray}
At zeroth and first order in this expansion we obtain the Einstein-Hilbert Lagrangian of GR with an effective cosmological constant term $\Lambda_{eff}= (\lambda-1)/(\kappa^2 \epsilon)$. Note that the next-to-leading order terms correspond to quadratic curvature corrections. Thus Born-Infeld gravity is an extension of GR that modifies its dynamics in regions of short lengths, as compared to the characteristic length-squared scale, $\epsilon$, present in the system.

As we are working in the Palatini approach, independent variations of the action (\ref{eq:EiBI}) with respect to metric and connection must be performed to obtain the field equations, which yields the two systems of equations
\begin{eqnarray}
\frac{\sqrt{ -q}}{\sqrt{-g}} q^{\mu\nu} - \lambda g^{\mu\nu}&=&-\kappa^2 \epsilon T^{\mu\nu} \label{eq:metricq} \\
\nabla_{\alpha} (\sqrt{-q} q^{\mu\nu})&=&0, \label{eq:connectionq}
\end{eqnarray}
where $q$ is the determinant of the (auxiliary) metric $q_{\mu\nu} \equiv g_{\mu\nu} + \epsilon R_{\mu\nu}$,  and $T_{\mu\nu}=-\frac{2}{\sqrt{-g}} \frac{\delta S_m}{\delta g^{\mu\nu}}$ is the stress-energy tensor of the matter. The convenience of introducing the metric $q_{\mu\nu}$ lies on the fact that Eq.(\ref{eq:connectionq}), which is fully equivalent to $\nabla_{\alpha} q^{\mu\nu}=0=\nabla_{\alpha} q_{\mu\nu}$, is solved as
 \begin{equation}
\Gamma^\lambda_{\mu\nu}= \frac{q^{\lambda\alpha}}{2}\left(\partial_\mu q_{\alpha\nu}+\partial_\nu q_{\alpha\mu}-\partial_\alpha q_{\mu\nu}\right),
\end{equation}
which means that the independent connection is given by the Christoffel symbols of $q_{\mu\nu}$. The latter is related to the physical metric $g_{\mu\nu}$ via the transformations
\begin{equation} \label{eq:q-g}
q_{\mu\nu}=g_{\mu\alpha }{\Omega^\alpha}_{\nu} ,
\end{equation}
where the matrix ${\Omega^\mu}_{\nu}$ is determined through Eq.(\ref{eq:metricq}) by the relation (hereafter a hat denotes a matrix)
\begin{equation} \label{eq:Omegadef}
| \hat{\Omega} | {(\Omega^{-1})^\mu}_{\nu}=\lambda {\delta^\mu}_{\nu} - \epsilon \kappa^2 {T^\mu}_{\nu} \ .
\end{equation}
It is important to realize that this equation provides a solution $\hat{\Omega} \equiv \hat{\Omega}({T^\mu}_{\nu})$ and thus the transformation (\ref{eq:q-g}) between $q_{\mu\nu}$ and $g_{\mu\nu}$ only depends on the matter sources. This, in turn, gives consistency to the introduction of the metric $q_{\mu\nu}$ to solve Eq.(\ref{eq:connectionq}). Now, using the relations (\ref{eq:q-g}) and (\ref{eq:Omegadef}) together with the definition of $q_{\mu\nu}$, the metric field equations (\ref{eq:metricq}) can be rewritten as
\begin{equation} \label{eq:Rmunuq}
{R_\mu}^{\nu}(q)=\frac{\kappa^2}{| \hat{\Omega} |^{1/2}} \left(L_G {\delta_\mu}^{\nu} + {T_\mu}^{\nu} \right),
\end{equation}
where the gravitational Born-Infeld Lagrangian, $L_G$, in Eq.(\ref{eq:EiBI}) is written as

\begin{equation}
L_G=\frac{ | \hat{\Omega} |^{1/2} - \lambda}{\epsilon \kappa^2}.
\end{equation}
Eqs.(\ref{eq:Rmunuq}) represent a system of Einstein-like, second-order field equations for the metric $q_{\mu\nu}$. All the contributions on the right-hand side of such equations are just functions of the matter and, as such, they can be collectively read as a modified stress-energy tensor. The algebraic and matter-dependent character of the transformations (\ref{eq:q-g}) guarantee the second-order character of the field equations for $g_{\mu\nu}$ as well. Note in passing that in vacuum, ${T_\mu}^{\nu}=0$, one has that $\hat{\Omega}=\hat{I}$ and $g_{\mu\nu}=q_{\mu\nu}$ (modulo a trivial re-scaling), and the field equations (\ref{eq:Rmunuq}) boil down to those of GR with a cosmological constant term. Thus, vacuum solutions are the same as those of GR and the theory is ghost-free\footnote{This seems to be a generic property of classical theories of gravity formulated in the Palatini approach \cite{Franca}.}, which means that only in presence of matter the non-trivial dynamics of Born-Infeld gravity is excited.

Regarding the matter sector, and to make contact with charged BTZ solutions \cite{Carlip:1995qv,MTZ}, let us consider a Maxwell electric field, whose action in a $2+1$ dimensional spacetime reads
\begin{equation} \label{eq:Maxwell}
S_m=-\frac{1}{16\pi l_P} \int d^3x \sqrt{-g} F_{\mu\nu}F^{\mu\nu},
\end{equation}
where $F_{\mu\nu}=\partial_{\mu}C_{\nu}-\partial_{\nu} C_{\mu}$ is the field strength tensor of the vector potential $C_{\mu}$. The matter field equations, $\nabla_{\mu} (\sqrt{-g}F^{\mu\nu})=0$, in a static spherically symmetric line element of the form
\begin{equation} \label{eq:lineg}
ds_g^2=-A(x)^2dt^2 + B(x)^2dr^2 +r^2(x) d\phi^2,
\end{equation}
and for a circularly symmetric field (whose components are $F_{0i} \neq 0$; $F_{ij}=0$, $i,j=1,2$), yield a solution that can be written as

\begin{equation}
X=\frac{Q^2}{r^2(x)},
\end{equation}
where $X \equiv -\frac{1}{2} F_{\mu\nu}F^{\mu\nu}$ does not depend explicitly on the metric functions $A(x)$ and $B(x)$, and $Q$ is an integration constant identified as the electric charge associated to a given solution. On the other hand, the stress-energy tensor for an electromagnetic field
\begin{equation}
{T_\mu}^{\nu}=-\frac{1}{4\pi l_P} \left({F_{\mu}}^{\alpha}{F_{\alpha}}^{\nu}-\frac{1}{4}{\delta_\mu}^{\nu} {F_\alpha}^{\beta}{F_{\beta}}^\alpha \right),
\end{equation}
in the $2+1$ dimensional circularly symmetric electrostatic spacetime (\ref{eq:lineg}) takes the form
\begin{equation} \label{eq:Tmunu}
{T_\mu}^{\nu}=\frac{X}{8\pi l_P}
\left(
\begin{array}{cc}
-\hat{I}_{2\times 2}&  \hat{0}_{1 \times 2} \\
\hat{0}_{2 \times 1} & 1 \\
\end{array}
\right),
\end{equation}
where $\hat{I}$ and $\hat{0}$ are the identity and zero matrices, respectively. With this matter source, we find that introducing the ansatz
\be \label{eq:Omegaem}
\hat{\Omega}=
\left(
\begin{array}{cc}
\Omega_{+} \hat{I}_{2\times2} &  \hat{0}_{1 \times 2} \\
\hat{0}_{2 \times 1} & \Omega_{-}, \\
\end{array}
\right),
\en
consistency with the definition (\ref{eq:Omegadef}) and the expression (\ref{eq:Tmunu}) leads to
\bea
\Omega_{-}&=&(\lambda + \tilde{X})^{2} \label{eq:omega1} \\
\Omega_{+}&=&(\lambda - \tilde{X} )(\lambda + \tilde{X}), \label{eq:omega2}
\ena
where $\tilde{X} \equiv \frac{\epsilon \kappa^2}{8\pi l_P} X=\frac{\epsilon \kappa^2 Q^2}{8\pi l_P r^2}$ is a dimensionless quantity (note that the charge $Q$ is also dimensionless). With all these expressions, the field equations (\ref{eq:Rmunuq}) in the present case follow immediately after a bit of algebra
\be \label{eq:Rmunuex}
\epsilon {R^\mu}_{\nu}(q)=
\left(
\begin{array}{cc}
\left(\frac{\Omega_{+}-1}{\Omega_{+}} \right) \hat{I}_{2\times2} &  \hat{0}\\
\hat{0} &  \left(\frac{\Omega_{-}-1}{\Omega_{-}} \right), \\
\end{array}
\right) \ .
\en
Computing the components of the Ricci tensor and noting that the right-hand side implies ${R_x}^x-{R_t}^t=0$, the resulting relation allows us to write the line element for $q_{\mu\nu}$ under the standard Schwarzschild-like form
\begin{equation}
ds_q^2=-\tilde{A}(x)^2dt^2+\frac{dx^2}{\tilde{A}(x)^2}+x^2 d\phi^2.
\end{equation}
On the other hand, the component ${R^\phi}_\phi$ of the field equations (\ref{eq:Rmunuex}) leads to the following equation (here a subindex means a derivative)
\begin{equation} \label{eq:Ax}
-\frac{[\tilde{A}(x)^2]_x}{x}=\frac{1}{\epsilon} \left(\frac{\Omega_{-}-1}{\Omega_{-}} \right),
\end{equation}
for the only independent metric component, $\tilde{A}(x)$. Using the relation (which follows from the transformations (\ref{eq:q-g}) in the angular sector)
\begin{equation} \label{eq:xr}
x^2=\Omega_- r^2,
\end{equation}
which by simple derivation yields

\begin{equation} \label{eq:dxdr}
dx=\frac{\Omega_+}{\Omega_-^{1/2}}dr,
\end{equation}
we can rewrite (\ref{eq:Ax}) as
\begin{equation}
-{[\tilde{A}(x)^2]_r}=\frac{r\Omega_+}{\epsilon} \left(\frac{\Omega_{-}-1}{\Omega_{-}} \right) \ ,
\end{equation}
and with a bit of algebra it can be put as
\begin{equation}
[\tilde{A}(x)^2]_r=-\frac{\lambda^2-1}{s |\epsilon|}r+s\frac{|\epsilon|}{r^3}Q^4-\left(\frac{Q^2}{\lambda}\right)\frac{d}{dr}\ln\left(\lambda r^2+s r_c^2\right).
\end{equation}
Upon integration, this leads to
\begin{equation} \label{eq:Atilde}
\tilde{A}(x)^2=k-\frac{\lambda^2-1}{2 s |\epsilon|}r^2-Q^2\left(\frac{s r_c^2}{2r^2}+\frac{1}{\lambda}\ln\left[\frac{\lambda r^2+s r_c^2}{r_0^2} \right]\right) \ ,
\end{equation}
where $k$ is an integration constant, $s$ denotes the sign of $\epsilon$, $r_c^2\equiv |\epsilon| \kappa^2Q^2/8\pi l_P=|\epsilon| Q^2$ sets the natural scale of the Born-Infeld gravitational corrections, and $r$ must be seen as a function of $x$ (due to Eq.(\ref{eq:xr})). The meaning of the constant $r_0$ in this expression comes from the fact that, due to the logarithm behavior of the electromagnetic field in $2+1$ dimensions, variation of the matter action (\ref{eq:Maxwell}) would yield a divergent surface boundary term at $r \to \infty$. Nonetheless, like in the standard BTZ case \cite{MTZ}, enclosing the system in a large circle of radius $r_0$ allows to cancel out such a boundary term, while the mass (which is related to the integration constant $k$, see below) becomes dependent on $r_0$. Though at the level of the field equations one could redefine constants to absorb $r_0$ into the mass parameter, we will leave $r_0$ explicit on the metric for convenience. It is also worth noting that the solution (\ref{eq:Atilde}) is invariant under $r\to -r$.

Now, using the transformations (\ref{eq:q-g}) with the expressions (\ref{eq:omega1}) and (\ref{eq:omega2}), the line element for $g_{\mu\nu}$ in Eq.(\ref{eq:lineg}) for these solutions can be written as
\begin{equation} \label{eq:linephys}
ds_g^2=-A(r)^2dt^2 + \frac{1}{A(r)^2} \left( \frac{dx}{\Omega_{+}}\right)^2 + r^2 (x) d \phi^2,
\end{equation}
with the definition $A^2 \equiv \frac{\tilde{A}^2}{\Omega_{+}}$ and where the explicit form of the relation $r=r(x)$ will be discussed in detail later. But before going into that we note that for a circularly symmetric metric in $2+1$ dimensions the constants of integration must be chosen in such a way that the constant term of the asymptotic far limit  of the metric represents the mass at infinity \cite{NED3D}. Thus, making series expansions of $A(r)$  for $r \gg r_c$ and identifying constants one gets $(k-\lambda^{-1}Q^2 \ln[\lambda])/\lambda^2=-M$. With this identification, and rearranging terms in (\ref{eq:Atilde}) we can write
\begin{equation} \label{eq:Aw}
\tilde{A}(x)^2 =-\lambda^2 M- \frac{\lambda^2-1}{2s|\epsilon|} r^2 -Q^2\left( \frac{sr_c^2}{2r^2}
+\frac{1}{\lambda}  \ln\left[\frac{r^2 + s r_c^2/\lambda}{r_0^2} \right]\right) \ .
\end{equation}
The far behavior of the metric function $A(r)= \frac{\tilde{A}}{\Omega_{+}}$, which coincides with the limit $|\epsilon|\to 0$, can be read directly from Eq.(\ref{eq:Aw}) because in this limit $\Omega_+=\lambda^2-(r_c/r)^4\approx \lambda^2$. The result is
\begin{equation} \label{eq:Arfar}
A(r)^2 \simeq -M -\frac{\lambda^2-1}{2\lambda^2 s |\epsilon| }r^2 - \frac{2Q^2}{\lambda^3} \ln\left[\frac{r}{r_0}\right] + O\left(\frac{\epsilon}{r^2}\right).
\end{equation}
In this limit, the leading order terms (for asymptotically AdS solutions) coincide with the solution of a $2+1$ electromagnetic field in the context of GR  \cite{Carlip:1995qv}
\begin{equation}\label{eq:AGR}
A_{GR}(r)^2=-M+\frac{r^2}{l^2_\Lambda}-2Q^2\ln\left[\frac{r}{r_0}\right],
\end{equation}
in agreement with the fact that Born-Infeld gravity reduces to GR in that limit. Comparison of (\ref{eq:Arfar}) with (\ref{eq:AGR}) allows us to identify $\tilde{\Lambda}_{eff} \equiv -1/l_\Lambda^2=\frac{\lambda^2-1}{2\lambda^2 s |\epsilon| }$ as the effective cosmological constant of the theory, which we will assume negative from now on. On the other hand, recalling that $|\epsilon|$ is a very small length scale squared, it follows that $\lambda$ must be extremely close to unity. This is something we will assume in practice, even if we keep $\lambda$ explicit everywhere. For AdS spaces, which are the ones we are most interested in, we must have $\lambda>1$ when $s=-1$ and $\lambda<1$ if $s=+1$.

From the far limit we thus see that only as one approaches the central region of these solutions do modifications to the solutions of GR appear. It is thus necessary to understand in detail those modifications.

\section{Analysis of the solutions} \label{sec:III}

\subsection{Wormhole structures}

With the expressions provided so far, the relation (\ref{eq:xr}) between $x$ and $r$ can be explicitly written as
\begin{equation}\label{eq:x(r)}
x^2=r^2\left(\lambda+s\frac{r_c^2}{r^2}\right)^2 \ ,
\end{equation}
which can be inverted to yield the result
\begin{equation} \label{eq:r(x)}
|r(x)|=\frac{|x|\pm \sqrt{|x|^2-4s\lambda r_c^2}}{2\lambda} \ ,
\end{equation}
where the plus sign  in front of the square root should be chosen in order to recover the correct asymptotic limit, $|r(x)| \simeq |x|$. The modulus is necessary to emphasize the invariance of the metric under the transformations $r\to -r$ and $x\to -x$.  In order to understand the behavior of the radial function $r(x)$ in (\ref{eq:r(x)}), let us split the discussion depending on the value of $s$ which, recall, corresponds to the sign of $\epsilon$.

\subsubsection{Case $s=-1$}

\begin{figure}[h]
\begin{center}
\includegraphics[width=1\textwidth]{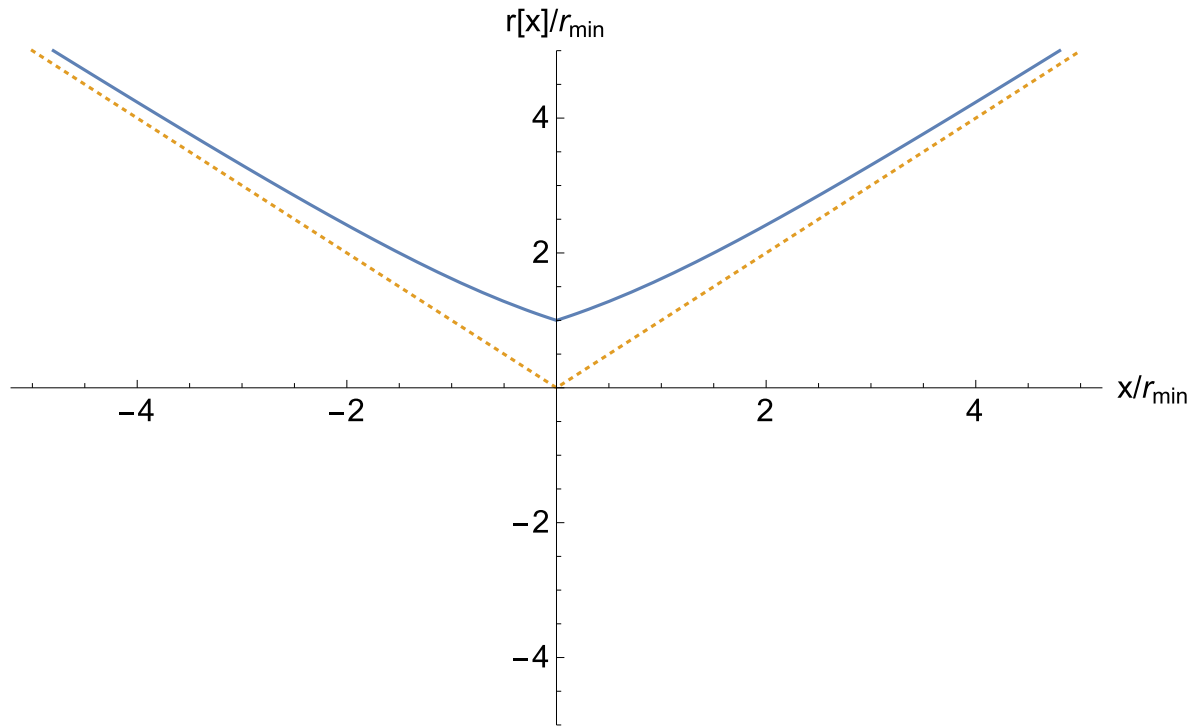}
\caption{ Representation of $r(x)$ for $s=-1$ (solid curve). Both axes are expressed in units of $r_{min}=r_c/\lambda^{1/2}$. Dashed lines represent the asymptotic GR behavior $r(x)=|x|$.  }\label{fig:r(x)}
\end{center}
\end{figure}

If $s=-1$ then $r(x)$ reaches a minimum of magnitude
\begin{equation} \label{eq:rminm}
r_{min}=\frac{r_c}{\lambda^{1/2}},
\end{equation}
at $x=0$ (see Fig.\ref{fig:r(x)}). This puts forward the existence of a minimum circumference of length $L_{min}=2\pi r_{min}$, which precludes the localization of the charges that generate the field at the center $r=0$. Rather than as a problem, this should be seen as a virtue of the theory. In fact, by admitting that $x$ is defined over the whole real line, the function $r(x)$ can be extended\footnote{Note, in this sense, that only $|x|$ enters into the definition of $r(x)$.} also to negative values of $x$ and one can interpret the electric charge of the field as a topological flux through a wormhole (see \cite{Visser} for a full account on wormhole physics), with $x=0$ representing the location of the wormhole throat\footnote{This clarifies why in the derivation of the field equations we have used the radial variable $x$ instead of $r(x)$. The reason is that $r$ is only a valid radial coordinate in those intervals where it is a strictly monotonic function.  Thus, one would need two charts in terms of $r(x)$ to cover the entire wormhole geometry, but a single one in terms of $x$.
}.
According to Gauss's theorem, the electric charge can be determined by the flux of electric lines flowing through the (circularly symmetric) wormhole throat as $Q=\int_{S^1} *F$, where $*F$ is Hodge dual of the field strength tensor. The possibility of defining a charge without making use of charge densities is at the heart of the charge-without-charge mechanism proposed by Misner and Wheeler long ago and on the introduction of the concept of \emph{geon} \cite{Wheeler}. Geometries with similar properties in four and higher dimensional Born-Infeld gravity with $s=-1$ have also been found recently \cite{ors,orhigher}, where these wormholes were identified as geons.

\subsubsection{Case $s=+1$}

\begin{figure}[h]
\begin{center}
\includegraphics[width=1.0\textwidth]{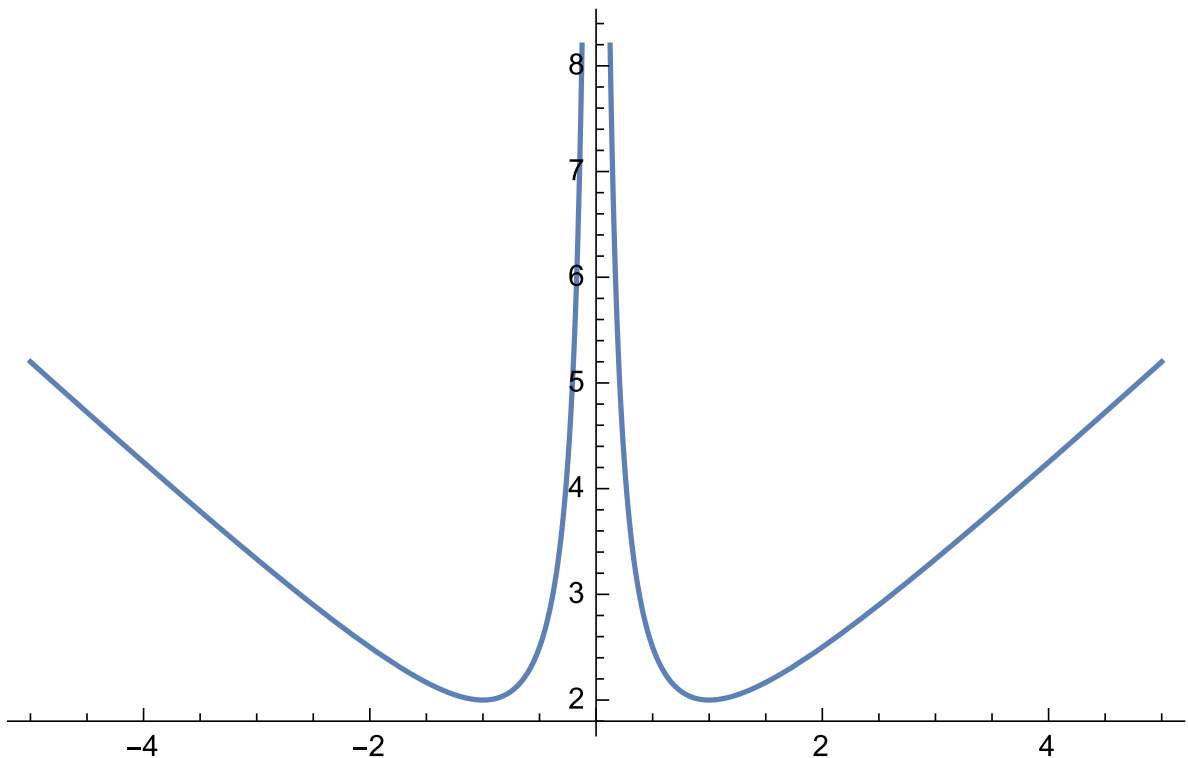}
\caption{ Representation of $|x(r)|$ for $s=+1$ (solid curve). Both axes are expressed in units of $r_{min}=r_c/\lambda^{1/2}$. Given that the metric is invariant under $r\to -r$, the two minima can be joined in a continuous way by identifying the points $r/r_{min}=+1$ and $r/r_{min}=-1$.   }\label{fig:x(r)}
\end{center}
\end{figure}

When $s=1$, then it is $|x(r)|$ that has a minimum of magnitude
\begin{equation}
|x_{min}|=2\lambda^{1/2}r_c \ ,
\end{equation}
which occurs at
(see Fig.\ref{fig:x(r)})
\begin{equation} \label{eq:rminp}
|r(x_{min})|= \frac{r_c}{\lambda^{1/2}}.
\end{equation}
Interestingly, this value of $r(x_{min})$ coincides with the minimum radius $r_{min}$ of the $s=-1$ case. The existence of this minimum for $x(r)$ indicates that $x$ is not a good coordinate to cover the whole geometry, in much the same way as $r$ was not appropriate for $s=-1$. The fact that $r(x_{\min})$ is not zero at the minimum of $x$ (see Fig.\ref{fig:x(r)}) also implies that extending $r$ to the negative axis cannot solve the problem either. In order to gain some insight on how to proceed, we find it useful to explore how this minimum is reached. In the $s=-1$ case  one finds that $r_-(x)\approx r_{min}+|x|/2\lambda$, whereas for $s=+1$ one gets  $r_+(x)\approx r_{min}+\sqrt{r_{min}(|x|-|x_{\min}|)}/\lambda^{3/4}$. This naturally motivates the introduction of a new coordinate $y$ (defined over the whole real line) in the form $|x|=|x_{min}|+y^2/(4\lambda r_{min})$ such that $r_+(y)\approx r_{min}+|y|/2\lambda$ takes the same functional dependence near its minimum as in the $s=-1$ case. This change in the radial coordinate implies the identification of the two minima of $|x(r)|$, which are located at $r=\pm r_{min}$ (see Fig.\ref{fig:x(r)}). Given that the metric is invariant under the change $r\to -r$, this identification does not imply any discontinuity in the metric. In fact, this transformation is similar in spirit to that used by Einstein and Rosen \cite{ER1935} in their construction of a particle model using the Schwarzschild geometry (the so-called {\it Einstein-Rosen bridge}). The resulting geometry is thus similar to that found in the case $s=-1$ and can also be interpreted as a wormhole generated by a topological electric flux.

\begin{figure}[h]
\begin{center}
\includegraphics[width=1.0\textwidth]{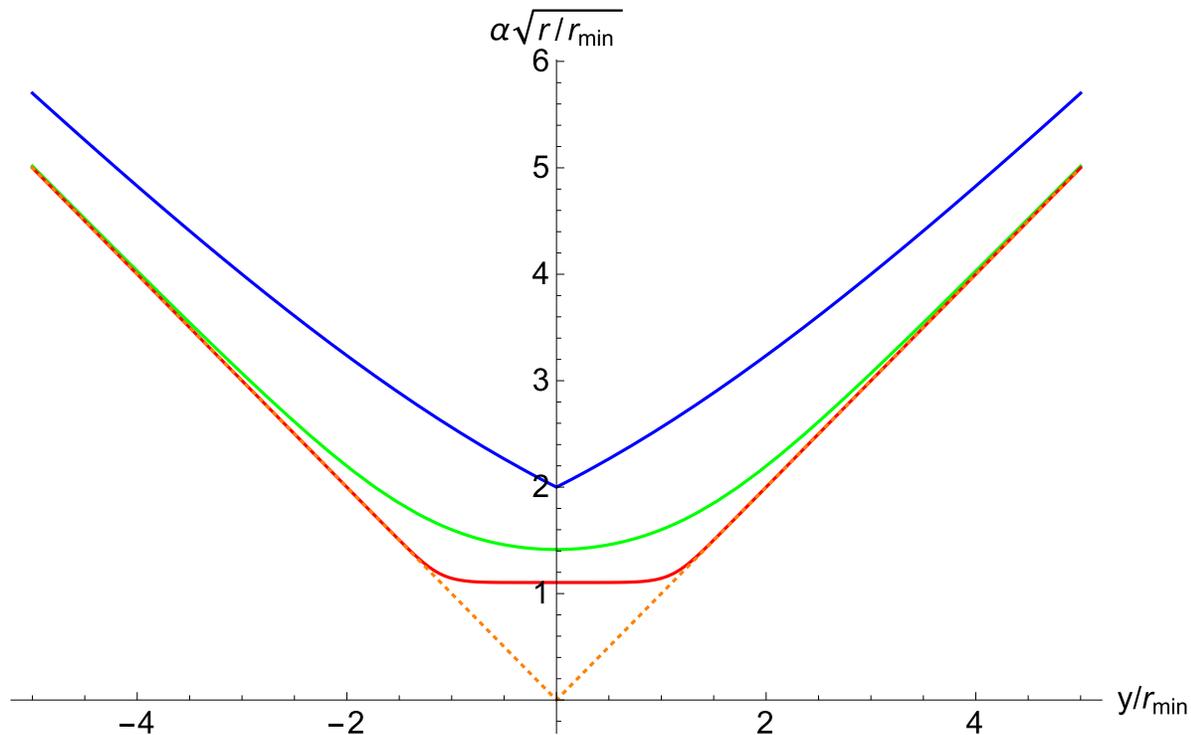}
\caption{ Representation of $\alpha \sqrt{r(y)/r_c}$ for $s=1$, with $\alpha=(4\lambda^{1/2})^{1/(2d-4)}$. Both axes are expressed in units of $r_{min}$ (and $\lambda=1$ for simplicity). The upper (blue) line represents the $d=3$ case, the middle line (green) is $d=4$, and the lower (red) curve is $d=9$. Dots represent the asymptotes $\pm y$. }\label{fig:WHs+1}
\end{center}
\end{figure}

It should be noted that this procedure is also valid in higher dimensions, where the relation (\ref{eq:x(r)}) between $x$ and $r$ takes the form
\begin{equation}
x^2=r^2\left(\lambda+s\left[\frac{r_c}{r}\right]^{2(d-2)}\right)^{\frac{2}{d-2}} \ ,
\end{equation}
which can be rewritten as
\begin{equation}
|x|^{2(d-2)}=|r|^{2(d-2)}\left(\lambda+s\left[\frac{r_c}{r}\right]^{2(d-2)}\right)^2 \ .
\end{equation}
Dividing on both sides by $r_c^{2(d-2)}$ and defining $|X|\equiv (|x|/r_c)^{d-2}$ and $|Z|\equiv (|r|/r_c)^{d-2}$, the resulting equation becomes identical with (\ref{eq:x(r)}) and, therefore, its solutions and extrema in terms of $X$ and $Z$ coincide with those already discussed above. The $d$-dimensional results, therefore, follow from those presented here by just  replacing $|r| \to |r|^{d-2}$ and $|x| \to |x|^{d-2}$. In that situation, $|r_{min}/r_c|^{d-2}=1/\lambda^{1/2}$,  $|x_{min}/r_c|^{d-2}=2\lambda^{1/2}$, and the change of coordinates that identifies the two minima of the function $x(r)$ takes the form $|x/r_c|^{d-2}=|x_{min}/r_c|^{d-2}+(y^2/r_c)^{d-2}/4\lambda^{1/2}$, which leads to $|r/r_c|^{d-2}=|r_{min}/r_c|^{d-2}+|y/r_c|^{d-2}/2\lambda$. In Fig.\ref{fig:WHs+1} the function $r_+(y)$ is shown for different values of the dimension $d$. Note that because $r(y)^{d-2}$ goes asymptotically like $\sim y^{2(d-2)}$, Fig.\ref{fig:WHs+1} actually represents $\sqrt{r_+(y)}$ (multiplied by a normalization constant). The case $s=+1$, therefore, admits a geon-like/wormhole  interpretation in much the same way as the $s=-1$ case studied in \cite{orhigher}. It is important to note that the function $r(y)$ near its minimum is smooth for all values of $d>3$ (see Fig.\ref{fig:WHs+1} as an illustration). Only the case $d=3$ has a discontinuous derivative at the minimum (vertex) due to its linear dependence on the modulus of $y$.  Something similar occurs when $s=-1$ (see Fig.\ref{fig:r(x)}).

\subsection{Non-wormhole solution}

In the $s=+1$ case, one may also consider the solution in which $r$ is defined within the standard interval $r\in[0,\infty[$. The metric function $\tilde{A}(r)$ takes then the form
\begin{equation} \label{eq:A}
\tilde{A}(r)^2 =-\lambda^2 M+\frac{\lambda^2r^2}{l_\Lambda^2} -Q^2\left( \frac{sr_c^2}{2r^2}
+\frac{1}{\lambda}  \ln\left[\frac{r^2 + s r_c^2/\lambda}{r_0^2} \right]\right) \ ,
\end{equation}
and the behavior near the center is dominated by the {\it angular momentum term} $-\frac{Q^2r_c^2}{2r^2}$, which has the {\it wrong} sign as compared to the BTZ solution of GR.  Using the relation (\ref{eq:dxdr}), the line element  (\ref{eq:linephys}) takes the form
\begin{equation} \label{eq:linephys2}
ds_g^2=-A(r)^2dt^2 + \frac{1}{A(r)^2}  \frac{dr^2}{\Omega_{-}} + r^2d \phi^2 \ ,
\end{equation}
and near $r\to 0$ becomes (here $d\tilde{t}=q^2dt/2$)
\begin{equation} \label{eq:linephys3}
ds_g^2=\frac{2r^2}{Q^2r_c^2}\left(-d\tilde{t}^{2} +{dr^2} + \frac{Q^2r_c^2}{2}d \phi^2\right) \ ,
\end{equation}
which looks like a conformally Minkowskian sphere of area $A=2\pi r_c^2 Q^2$.

\subsection{Comparison to BTZ black hole with charge}

According to Carlip \cite{Carlip:1995qv}, the line element of the BTZ black hole with (negative) cosmological constant $\Lambda<0$ ($l_{\Lambda}^2>0$), mass $M$, charge $Q$, and angular momentum $J$, can be written as
\begin{equation}
ds^2=-f(r)^2dt^2+\frac{1}{f(r)^2}dr^2+r^2\left(d\phi-\frac{J}{2r^2}dt\right)^2 \ ,
\end{equation}
where
\begin{equation}
f(r)^2=-M+\frac{r^2}{l_{\Lambda}^2}+\frac{J^2}{4r^2}-\frac{Q^2}{2}\ln \frac{r}{r_0} \ .
\end{equation}
The functional resemblance between this expression and our results is remarkable. Aside from the $J-$dependence of the crossed term $d\phi dt$, the time component is essentially the same up to a redefinition of constants. In fact, the $J^2$ term is now given by $J^2\to -2s Q^2r_c^2$, which is related to the $|\epsilon|$ parameter of Born-Infeld gravity and the electric charge via $r_c^2=|\epsilon|Q^2$. Its sign can also be controlled, not being fixed to be positive. This puts forward that even without angular momentum, our solutions can have event horizons (see Sec.\ref{sec:Hor} below).

The logarithmic dependence is also very interesting and implies the existence of a limiting value for the radial coordinate, which cannot drop below the scale $r_{min}=r_c/\lambda^{1/2}$, when $s=-1$. As pointed out above, this is a sign that a wormhole structure exists. Something similar can also be inferred for the $s=+1$ case due to the behavior of the function $x^2(r)$, as has been already discussed. It is also worth mentioning that the correction $a\equiv sr_c^2/\lambda$ appearing in the logarithm is similar to some models studied in the context of GR \cite{Cataldo}. In those models, such corrections were introduced \emph{by hand}, i.e., by defining first the geometry and then driving the Einstein equations back in order to find the matter source supporting it. The resulting such source can be interpreted as a kind of non-linear electrodynamics model, with the peculiarity that the corresponding solutions yield finite curvature scalars everywhere (note that such models are known to be physically problematic, though \cite{Bronnikov}). In contrast, in our case the new form of the metric arises from an {\it a priori} defined action, i.e, not designed in an {\it ad hoc} manner.

\subsection{Curvature scalars}

Regarding the presence of curvature divergences for the solutions above, one may compute the Kretschmann scalar, $K \equiv {R^\alpha}_{\mu\alpha\beta}{R_\alpha}^{\mu\alpha\beta}$, for both cases $s= \pm 1$. Though long expressions are obtained, one can expand them in the relevant regions. Taking $\lambda=1$ for simplicity, for $r \gg r_{min}$ one gets the result

\begin{equation}
K(r \gg r_{min}) \simeq \frac{12Q^4}{r^4} \mp s  \frac{60 r_c^2Q^4}{r^6} + O\left(\frac{1}{r^{8}}\right),
\end{equation}
where the first term corresponds to the GR behavior. This is the same result as one would obtain when the limit $|\epsilon| \ll 1$ is considered. In the wormhole cases, when the Kretschmann is expanded around the minimum of the radial coordinate, $r_{min}$, one gets a divergence to leading order
\begin{equation}
K(r \to r_{min}) \sim \frac{\alpha(M,Q,\lambda,s,|\epsilon|,r_0)}{(r-r_{min})^4} + O\left(\frac{1}{(r-r_{min})^3}\right),
\end{equation}
where $\alpha(M,Q,\lambda,s, |\epsilon|,r_0)$ is a constant whose particular value is not relevant for this discussion. Thus, this divergence is of the same magnitude as in the GR case, though it is now displaced to the circumference of radius $r_{min}$. In the non-wormhole solution of the $s=+1$ case, the Kretschmann near the origin $r\to 0$ diverges as $\sim Q^4r_c^4/r^8$. This divergence is reached after crossing the one located at $r=r_{\min}$, where the function $\Omega_+$ vanishes. To see what this implies upon the regularity of the corresponding spacetimes we will have to study the geodesic structure of the solutions, as we shall see in Sec.\ref{sec:Geo}. But before going that way let us have a look at the horizons of these solutions.

\subsection{Horizons} \label{sec:Hor}

\begin{figure}[h]
\begin{center}
\includegraphics[width=1.0\textwidth]{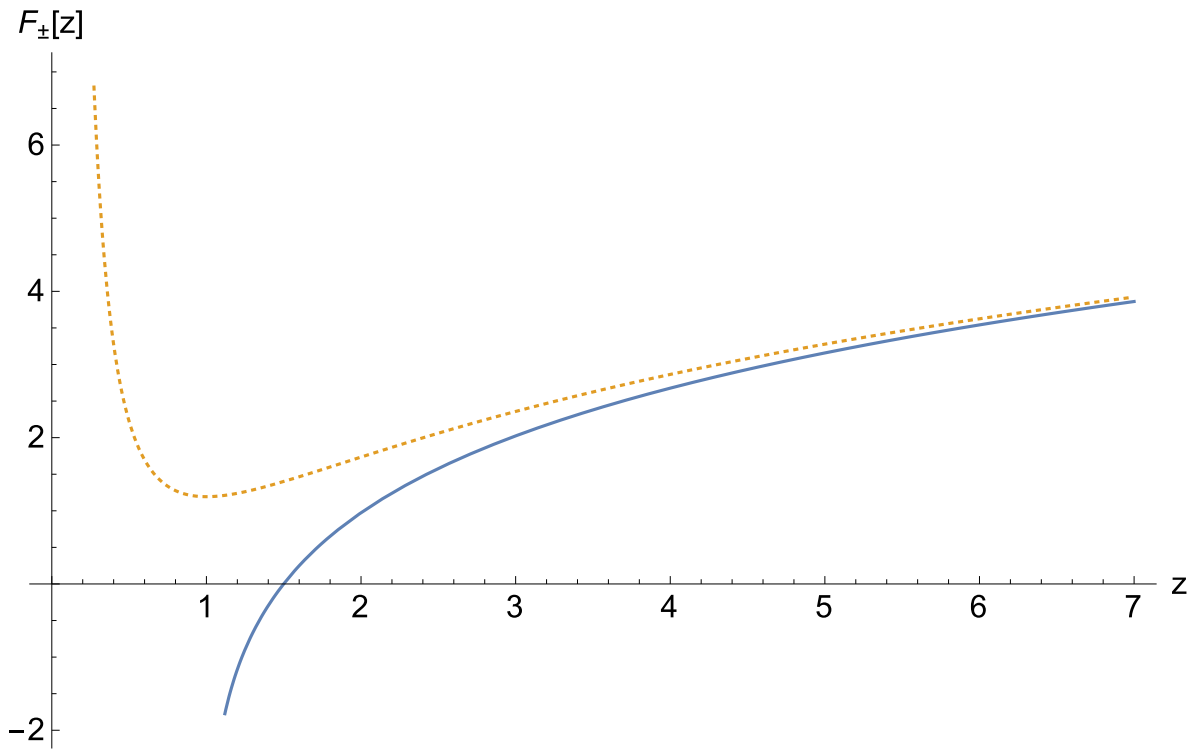}
\caption{Representation of $F_s(z)$. The solid line represents $s=-1$ and the dotted one $s=+1$.  }\label{fig:Fpm}
\end{center}
\end{figure}

For the line element (\ref{eq:linephys}), the horizons are given simply by the zeros of $A(r)^2=\frac{\tilde{A}(r)^2}{\Omega_{+}}$. The presence of the logarithm makes it impossible to find closed analytical expressions for the values of $r$ at which the horizons are located. For this reason, we prefer to adopt a relaxed perspective and provide a qualitative discussion of the number and location of the possible horizons. We begin by introducing a dimensionless radial variable $z=r/r_{min}$ in terms of which $A(r)^2$ becomes
\begin{eqnarray}
{A}(r)^2 &=&\frac{Q^2/\lambda^2}{1-\frac{1}{z^4}}\left[-\frac{M_{eff}^2}{Q^2} -\left( \frac{s\lambda}{2z^2} +\frac{1}{\lambda}  \ln\left[{z^2 + s }\right]\right)\nonumber \right. \\ &+& \left. \frac{\lambda^2r_{min}^2}{Q^2l_\Lambda^2}z^2 \right] \ ,
\end{eqnarray}
where $M_{eff}\equiv \lambda^2 M+\frac{2Q^2}{\lambda}\ln[r_{min}/r_0]$. For the sake of clarity, we will assume $\lambda\to 1$ everywhere except in the definition of $l_\Lambda^2$. So, in practice, we will be dealing with
\begin{eqnarray}
{A}(r)^2 &=&\frac{Q^2}{1-\frac{1}{z^4}}\left[-\frac{M_{eff}^2}{Q^2} -\left( \frac{s}{2z^2} + \ln\left[{z^2 + s }\right]\right)\nonumber \right. \\ &+& \left. \frac{r_{min}^2}{Q^2l_\Lambda^2}z^2 \right] \ .
\end{eqnarray}
Moreover, we will assume that $l_\Lambda^2\gg r_{min}^2$, in such a way that the term $\frac{r_{min}^2}{Q^2l_\Lambda^2}z^2$ is negligible as the region $z\to 1$ is approached. The presence or not of horizons is thus crucially determined by the behavior of the function $F_{s}(z)\equiv \left( \frac{s}{2z^2} + \ln\left[{z^2 + s }\right]\right)$ around $z\to 1$ (see Fig.\ref{fig:Fpm}).

\subsubsection{Case $s=-1$.}
When $s=-1$, we find that $\lim_{z\to 1^+}F_-(z)\to -\infty$.  As a result, for any finite value of $M_{eff}/Q^2$ (regardless of its sign) we have $A^2>0$ there. For $z\simeq 1.5$,  we have $F_-(z)>0$ and given that  $\frac{r_{min}^2}{Q^2l_\Lambda^2}z^2$ is negligible, $A^2$ must vanish at some $z\geq 1.5$ if $M_{eff}/Q^2>0$ or at  $1<z\leq 1.5$ if $M_{eff}/Q^2<0$. This sets the location of an inner horizon $z_{in}$. An outer horizon must arise at some $z_{out}>z_{in}$ when the $\frac{r_{min}^2}{Q^2l_\Lambda^2}z^2$ term grows enough to compensate the negative contribution of $-M^2_{eff}/Q^2-F_-(z)$. Depending on the particular values of the parameters, configurations with one (degenerate) horizon and with no horizons are also possible.

\subsubsection{Case $s=+1$ with wormhole at $z=1$.}

When $s=+1$, we find that $\lim_{z\to 1^+}F_+(z)\to 1/2+\ln 2\approx 1.19315 \equiv \zeta$. This implies that if $M_{eff}/Q^2>-\zeta$ then $\lim_{z\to 1^+}A^2<0$, which implies that this region is hidden behind an event horizon.  At larger values of $z$ there must be a point at which the logarithmic growth of $F_+(z)$ and the AdS term $\sim z^2$ meet to set the location of the horizon. These configurations, therefore, can be regarded as Schwarzschild-like and are not found in the case of standard BTZ solutions of GR. On the other hand, if $M_{eff}/Q^2<-\zeta$, we have that $\lim_{z\to 1^+}A^2>0$. The growth of $F_+(z)$ must lead to an inner horizon at some $z>1$. When the $z^2$ term dominates, an outer horizon emerges defining the region where $A^2(z)>0$ again. Obviously, this Reissner-Nordstr\"{o}m-like configuration may also have cases with a single (degenerate) horizon and with no horizons, depending on the specific values of the parameters.

\subsubsection{Case $s=+1$ without wormhole at $z=1$.}

The previous discussion also applies to this one up to $z\to1^+$. As $z=1$ is crossed, the $(1-1/z^4)$ term in the denominator of $A^2$ implies a divergence and a change of sign. This change of sign is not due to a zero in $A^2$ and, therefore, $z=1$ cannot be regarded as a typical horizon. For $0\leq z<1^-$ we need to note that $A^2>0$ everywhere if  $M_{eff}/q^2>-\zeta$, whereas for $M_{eff}/q^2<-\zeta$ the transit from $1^+$ to $1^-$ involves a change of sign that makes $A^2<0$. As $z\to 0$, $F_+(z)$ grows rapidly  and $A^2$ changes sign again, defining a new inner horizon. \\

\subsection{Thermodynamics}

From the horizon structure, and the presence of event horizons in some of solutions above, one could go further and obtain the thermodynamical behavior. The temperature for a circularly symmetric metric is defined and obtained in our case as \cite{Wald}
\begin{equation} \label{eq:T}
T=\frac{1}{4\pi} \lim_{r-r_h} \frac{\partial_r g_{tt}}{\sqrt{-g_{tt}g_{rr}}} = \frac{\Omega_{-}^{1/2}}{4\pi} \frac{d}{dr} \left(\frac{\tilde{A}(r)}{\Omega_{+}} \right) \Big\vert_{r=r_h}
\end{equation}
where $r_h$ denotes the black hole event horizon (for which $dg_{tt}/dr \vert_{r=r_h}>0$), and in the last equality we have used the relation of coordinates of Eq.(\ref{eq:dxdr}) to write the whole expression in the same coordinate system (this is fully equivalent to using the coordinate $x$). The discussion of the temperature is highly non-trivial due to the involved interplay between $M,Q,\lambda,s, |\epsilon|$, and the lack of analytic expressions for the horizon $r_h$ in terms of elementary functions. Nonetheless, it is easy to check that the temperature in (\ref{eq:T}) is consistent with the charged BTZ solution ($\epsilon \to 0$) and the vacuum one ($Q \to 0$), as follows from the asymptotic expansion (\ref{eq:Arfar}). It is also worth pointing out that the entropy of black holes in $2+1$ dimensions, $S=2\pi r_h$, does not necessarily holds in the case of extended theories of gravity. This has been explicitly verified in several such extensions in four spacetime dimensions \cite{Entropy}. The corresponding research on this issue in the case of $2+1$ dimensions has been scarcely explored and, to the best of our knowledge, it is not available for Born-Infeld gravity yet. As this issue goes beyond the scope of this work, we shall not discuss it further here.

\section{Geodesics} \label{sec:Geo}

In the original formulation of the singularity theorems \cite{Theorems} (see \cite{Senovilla} for a pedagogical discussion), the notion of \emph{geodesic completeness} is of central importance to determine whether a given spacetime is singular or not. In this sense, no reference to the possible divergence of curvature scalars is made at all, and the attention is focused on whether \emph{all} null and time-like geodesics can be extended to arbitrarily large values of their affine parameter or not. The physical content of this statement is that, being time-like and null geodesics attached to the free falling evolution of physical observers and the transmission of information, respectively, nothing should be allowed to suddenly disappear from the manifold or emerge out of nowhere. In this sense, it is worth pointing out that the widespread identification found in the literature between spacetime singularities and curvature divergences comes from the fact that, in most cases of physical interest, those spacetimes having divergent curvature scalars are also geodesically incomplete (see e.g. \cite{Ansoldi} for a review of regular solutions in the context of GR). Using geodesic completeness, under the assumptions of the existence of trapped surfaces, the null congruence condition (equivalent to the null energy condition via Einstein's equations) and global hiperbolicity, the unavoidable existence of spacetime singularities within GR is proved. Here we explore whether our solutions represent singular spacetimes or not by studying their geodesic structure.

To investigate this issue in detail, we note that for static spacetimes with line element $ds^2=-C(x)dt^2+\frac{1}{B(x)}dx^2+r^2(x)d\phi^2$, by spherical symmetry one can take the movement to occur in the equatorial plane $\phi=\pi/2$ without loss of generality, and geodesics are described by the following equation \cite{Chandra}
\begin{equation}
\frac{C(x)}{B(x)}\left(\frac{dx}{d\sigma}\right)^2=E^2-C(x)\left(\frac{L^2}{r^2}-k\right) \ ,
\end{equation}
where $\sigma$ is the affine parameter and $k=-1,0,1$ for time-like, null-like, and space-like geodesics, respectively. For time-like geodesics, the conserved quantities $E=Cdt/d\sigma$ and $L=r^2(x)d\phi/d\sigma$ carry the meaning of the total energy per unit mass and angular momentum per unit mass, respectively, while for null geodesics $L/E$ can be identified as the impact parameter. In our case, for the line element (\ref{eq:linephys}) this equation can also be written using the relation between coordinates (\ref{eq:xr})  as
\begin{equation}\label{eq:geodesics}
\frac{1}{\Omega_-}\left(\frac{dr}{d\sigma}\right)^2=E^2-A^2\left(\frac{L^2}{r^2}-k\right) \ .
\end{equation}
As null and time-like geodesics are associated to the transmission of information and free-falling evolution of idealized observers, respectively (while no observer or particle is known to be able to move in a space-like geodesic), in what follows we shall split our analysis into two different cases, and study the geodesic evolution of the wormhole and non-wormhole families.

\subsection{Radial null geodesics}

For null radial geodesics, $k=0=L$, the above equation leads simply to
\begin{equation} \label{eq:nullgeo}
\frac{1}{\left(\lambda+\frac{sr_c^2}{r^2}\right)}\frac{dr}{d\sigma}=\pm E \ ,
\end{equation}
where the plus/minus sign represents outgoing/ingoing geodesics. Care should be taken when the geodesics cross the wormhole throat $r_{min}$ because then the signs must be reversed to account for the change of orientation of the normal to the circularly symmetric wormhole throat. For $s=-1$, the integration leads to
\begin{equation}
\frac{r}{r_{min}}+\frac{1}{2}\ln\left(\frac{r-r_{min}}{r+r_{min}}\right)=\pm E\lambda(\sigma-\sigma_0) \ .
\end{equation}
where $\sigma_0$ is an integration constant.

\begin{figure}[h]
\begin{center}
\includegraphics[width=1.0\textwidth]{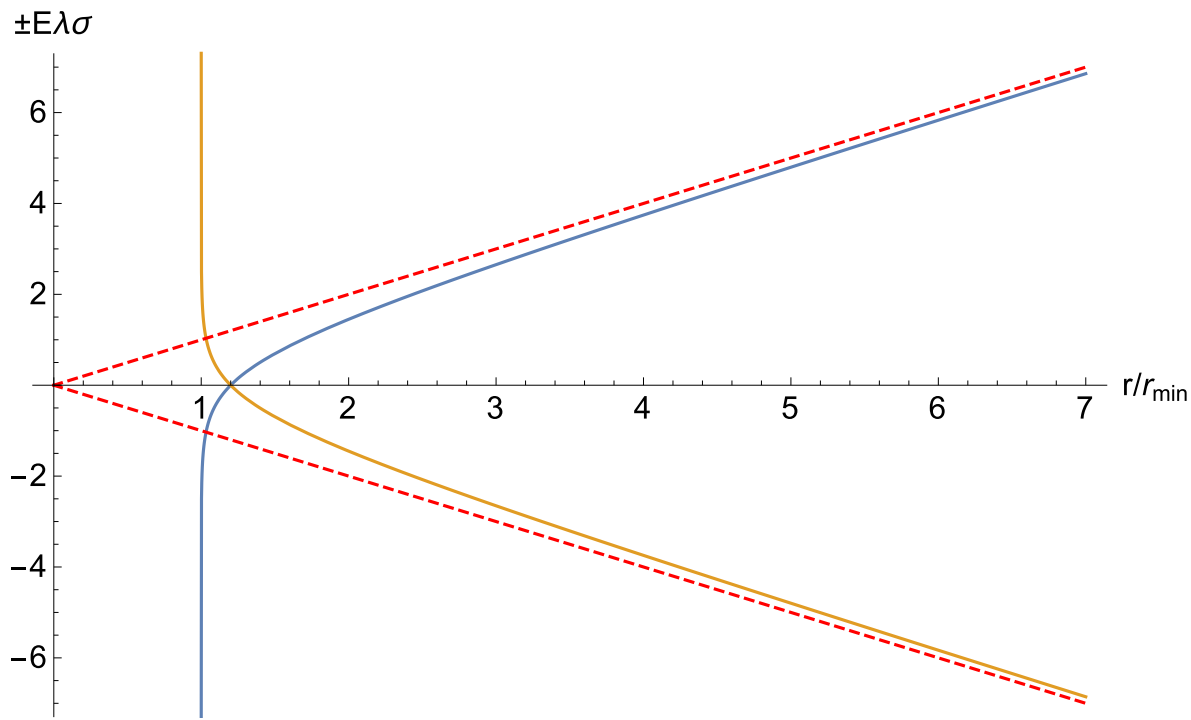}
\caption{Affine parameter $\sigma(x)$ for ingoing and outgoing radial null geodesics for $s=-1$. For $r\gg r_{min}$ one recovers the GR behavior (dashed straight lines), $r\approx \pm \sigma$, while when $r\to r_{min}$  one has $\sigma\to \mp \infty $ (solid blue and orange curves) and the geodesics become complete.  In this plot $E=\lambda=1$.}\label{fig:Nullsp1WH.eps}
\end{center}
\end{figure}
For $r\gg r_{min}$ one recovers the usual behavior of GR, $r\approx \pm \sigma$. Now, the presence of a minimum in the radial coordinate induces the behavior $\sigma\to \mp \infty $ as $r\to r_{min}$. This shows that such geodesics are complete (see Fig.\ref{fig:Nullsp1WH.eps}). This result holds true for all the spectrum of mass and charge of the solutions, i.e. regardless of the existence of horizons, and is also insensitive to the presence of a cosmological constant. Let us point out that the fact that the affine parameter for null geodesics diverges at the location of the wormhole throat  means that the wormhole actually lies on the future (or past) boundary of the geometry and, therefore, cannot be reached in any finite affine time. This mechanism of resolution of spacetime singularities has been already observed in certain four dimensional $f(R)$ models \cite{bcor}.

\begin{figure}[h]\begin{center}
\includegraphics[width=1.0\textwidth]{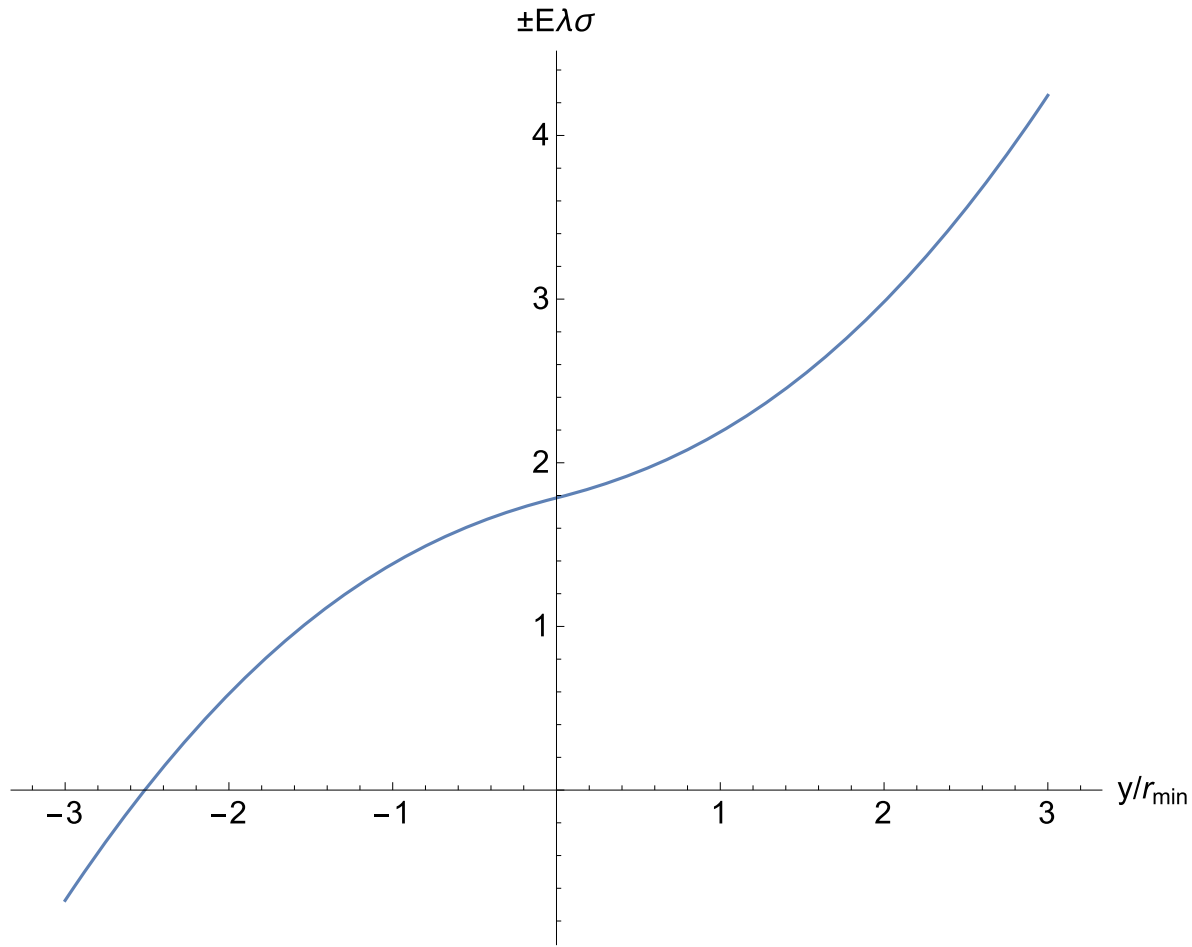}
\caption{Affine parameter $\sigma(r[y])$ for radial null geodesics in the case $s=+1$ with wormhole configuration. As opposed to the $s=-1$ cases, these geodesics reach the wormhole throat $r_{min}$ in a finite affine time but can be extended beyond this point, which guarantees their completeness. In this plot $E=\lambda=1$.}\label{fig:Nullsp1WH}
\end{center}
\end{figure}

When $s=+1$ and the wormhole branch is considered,  the solution  to Eq.(\ref{eq:nullgeo}) becomes
\begin{equation}
\pm E\lambda(\sigma-\sigma_0)=\left\{\begin{array}{lr} \frac{r}{r_{\min}}-\arctan\left(\frac{r}{r_{min}}\right) +\frac{\pi}{2} & if \hspace{0.2cm}  y>0 \\  \\
\alpha- \frac{r}{r_{\min}}+\arctan\left(\frac{r}{r_{min}}\right)+\frac{\pi}{2} &  if \hspace{0.2cm}   y<0 \end{array}\right.
\end{equation}
where $\alpha=2\left(1-\frac{\pi}{4}\right)$ is a constant and $r=r(x)$ is given by (\ref{eq:r(x)}) with $|x|=|x_{\min}|+y^2/4\lambda r_{min}$. The factor $\pi/2$ has been added so as to get the asymptotic GR limit, $r\approx \pm E\lambda(\sigma-\sigma_0)$, when $y\to +\infty$. As depicted in Fig.\ref{fig:Nullsp1WH}, these geodesics reach the wormhole throat $r_{min}$ in a finite affine time. However, it is also shown that they can be extended beyond this point to arbitrarily large values of the affine parameter, which implies that they are also complete for all the spectrum of mass, charge, energy and cosmological constant term. As such, they are also insensitive to the presence of curvature divergences at the wormhole throat, whose meaning and implications should be separately analyzed (see \cite{ors16} for some insights on the four dimensional case).

In the $s=+1$ case without wormhole structure, radial null geodesics are just given by $\pm E\lambda(\sigma-\sigma_0)= \frac{r}{r_{\min}}-\arctan\left(\frac{r}{r_{min}}\right)+\frac{\pi}{2}$. Like in the previous case, for $r\gg r_{min}$ we recover the GR limit. Given that the causal structure as one approaches the center  is (conformally) Minkowskian, see Eq.(\ref{eq:linephys3}), nothing prevents that incoming geodesics reach the region $r=0$ in a finite affine time. The situation is thus similar to what occurs in the case of the four-dimensional Reissner-Nordstr\"{o}m black hole and, therefore, this solution can be regarded as singular\footnote{We note that in the Reissner-Nordstr\"{o}m case, the point-like sources that generate the electric field are not specified and, for this reason, the field equations are not solved at $r=0$. This implies that $r=0$ represents a limiting boundary of the spacetime. The fact that some geodesics can reach it in a finite affine time legitimate us to treat this solution as singular.}.

\subsection{Timelike and nonradial geodesics}

For time-like, $k=-1$, and non-radial geodesics, $L\neq 0$, the behavior depends on the parameters of the particular solution considered. Since the GR limit is quickly recovered just a few $r_{min}$ units away from $r_{min}$, as the null geodesics show, we will consider only those geodesics that are able to approach to $r_{min}$.

In the $s=-1$ case, we saw that $A^2>0$ and diverges as $r\to r_{min}$. As a result, the right-hand side of (\ref{eq:geodesics}) necessarily vanishes at some $r>r_{min}$, which implies a minimum in the function $r(\sigma)$. This means that all time-like and $L\neq 0$ geodesics bounce before reaching $r_{min}$. A similar behavior appears when $s=+1$ and $M_{eff}/Q^2<-\zeta$. However, for $M_{eff}/Q^2>-\zeta$, we have  $A^2<0$ (and divergent as $r\to r_{min}^+$), which allows to reach $r=r_{min}$.
If no wormhole structure is assumed, just after crossing $r_{min}$, in the region $r_{min}^-$ the $\Omega_+$ term in the denominator of $A^2$ changes sign and turns the infinite attractive potential of the geodesic equation into an infinite repulsive barrier. The right-hand side of (\ref{eq:geodesics}) then becomes negative for any finite energy and the particle cannot get into that region. It thus follows that geodesics with $k=-1$ and/or $L\neq 0$ cannot be extended into the region $r<r_{min}$ for these non-wormhole solutions. Such geodesics are incomplete, which provides further evidence to support that  these solutions represent singular spacetimes. If on the contrary, we consider the wormhole extension of these solutions, the geodesics that propagate in the $r>r_{min}$ (with coordinate $y>0$) black hole region ($A^2<0$) must fall towards $r\to r_{min}$ and then bounce into the other region $r>r_{min}$ (with $y<0$), which acts as a white hole due to the change of orientation in $dr/dy$ that occurs after the minimum at $r_{min}$ (see \cite{ors15v} for a discussion of this type of solutions in four dimensions).

It thus follows that all solutions with wormhole structure are geodesically complete regardless of the sign of $\epsilon$. The solution in which $r$ has been allowed to reach the center has null incomplete geodesics is also geodesically complete if $M_{eff}/Q^2<-\zeta$. For $M_{eff}/Q^2>-\zeta$, time-like and non-radial null geodesics are also incomplete.

\section{Summary and conclusions} \label{sec:V}

In this work we have considered the problem of electrically charged, nonrotating solutions of Born-Infeld gravity in $2+1$ dimensions with cosmological constant.  We found a solution for the metric components in closed, analytic form which recovers the GR case in the corresponding limit ($\epsilon\to 0$). This solution resembles the charged rotating BTZ solution of GR in that a new term induced by the Born-Infeld gravity corrections has the shape of a $J^2$ contribution. Beyond this resemblance, these solutions contain some nontrivial novelties such as an enlarged description in terms of (type and number of)  horizons and the presence of finite-size wormhole structures for all the spectrum of mass and charge. When the Born-Infeld gravity parameter $\epsilon$ vanishes, the wormhole throat closes and one recovers the standard point-like structure of GR. A (pathological) family of solutions without wormhole structure has also been found.

The most important result of this work is the verification that all the wormhole solutions found are geodesically complete and, therefore, can be regarded as nonsingular spacetimes. This result is independent of the existence or not of event horizons.  Curvature divergences generically arise at the wormhole throat but are not an obstacle for the completeness of geodesics. Therefore, even {\it naked divergences} should be regarded as physically admissible solutions.

Though, as is common in the literature, we have focused on the behavior of geodesic observers to characterize the singular/nonsingular character of the space-time, one should also consider the fate of accelerated observers. In this sense, if observers with finite proper acceleration were able to reach the boundaries of the space-time in a finite affine time, then it would be difficult to claim that such spaces are nonsingular. The non-traversable wormholes of the case $s=-1$ represent one such boundary. Preliminary results from \cite{WiP} indicate that the solutions found here are also safe for accelerated observers with bounded radial proper acceleration (see \cite{ao} for some recent results on uniformly accelerated observers in the literature). Thus, our claims based on geodesic completeness seem to be robust.

The solutions without wormhole structure have incomplete radial null geodesics for the same reasons as the Reissner-Nordstr\"{o}m solution in four dimensions, namely, because the limiting boundary $r=0$ is reached in a finite affine time. When the condition $M_{eff}/q^2<-\zeta$ is satisfied, where $M_{eff}\equiv \lambda^2 M+\frac{2q^2}{\lambda}\ln[r_{min}/r_0]$, then time-like and non-radial geodesics are reflected before reaching $r=r_{min}$. For $M_{eff}/q^2>-\zeta$, time-like and non-radial geodesics ($L\neq 0$) terminate at the circumference $r=r_{min}$ because causality prevents them from bouncing back to $r>r_{min}$ and an infinite barrier prohibits their extension below $r_{min}$.

It turns out that the replacement of the GR-point like singularity by a wormhole seems to be quite a generic prediction of extensions of GR formulated in the Palatini approach, where the ghost-free and second-order character of the field equations is not restricted to the case of Born-Infeld gravity, but also extends to the more standard scenario where curvature scalars are directly added to the gravitational action \cite{Torsion}. On the other hand, the result regarding wormholes has been observed in four spacetime dimensions both for $f(R)$ gravity \cite{bcor} and extensions containing Ricci-squared corrections \cite{Riccisquared}, in Born-Infeld gravity (with $s=-1$ \cite{ors}), as well as in higher-dimensional generalizations \cite{orhigher}. The results obtained in this paper are in agreement with the two different mechanisms for the resolution of spacetime singularities observed in those works: either the wormhole lies on the future (or past) boundary of the spacetime, since it is reached by null geodesics in an infinite affine time (this is what occurs in the $s=-1$ case, see Fig.\ref{fig:Nullsp1WH.eps}), or the wormhole can be reached in a finite affine time by null and time-like geodesics but these can be smoothly extended from the throat to arbitrarily large values of the affine parameter ($s=+1$ case, see Fig.\ref{fig:Nullsp1WH}). The latter solutions, which can be easily generalized beyond the $2+1$ dimensional case, represent a feature previously unnoticed in the literature of Born-Infeld gravity in four and higher dimensions.

To conclude, these novelties and the regular character (as given by geodesic completeness) of all the BTZ-type wormhole solutions for the whole spectrum of mass and charge in Born-Infeld gravity represent an interesting extension of the standard BTZ solution of GR. Subsequent analysis of these results might enlarge the thermodynamic and holographic applications of the BTZ-like solutions in $2+1$ spacetimes. The presence of curvature divergences at the wormhole throat also deserve further investigation to understand their implications. Work along these lines is currently underway.

\section*{Acknowledgments}

The work of D.B and L.L. is partially supported by the CNPq (Brazil) grants 455931/2014-3 and 306614-2014-6, and 307111/2013-0 and 447643/2014-2, respectively. G. J. O. is supported by a Ramon y Cajal contract and the Spanish grant FIS2014-57387-C3-1-P from MINECO.  D.R.G. is funded by the Funda\c{c}\~ao para a Ci\^encia e a Tecnologia (FCT, Portugal) postdoctoral fellowship No.~SFRH/BPD/102958/2014 and the FCT research grant UID/FIS/04434/2013. Support from the Consolider Program CPANPHY-1205388, the Severo Ochoa grant SEV-2014-0398, and the CNPq project No.301137/2014-5 is also acknowledged. This article is based upon work from COST Action CA15117, supported by COST (European Cooperation in Science and Technology).

\end{document}